\documentclass[aip,graphicx]{revtex4-1}
\usepackage{graphicx} 
\usepackage{epstopdf}

\draft 

\begin{document}


\title{Impact of resonator geometry and its coupling with ground plane on ultrathin metamaterial perfect absorbers} 



\author{Li Huang}
\thanks{These two authors contributed equally to this work}
\affiliation{Physics Department, Harbin Institute of Technology, Harbin, Heilongjiang 150001, China}

\author{Dibakar Roy Chowdhury}
\thanks{These two authors contributed equally to this work}
\affiliation{Center for Integrated Nanotechnologies, Los Alamos National Laboratory, Los Alamos, New Mexico 87545, USA}

\author{Suchitra Ramani}
\affiliation{Center for Integrated Nanotechnologies, Los Alamos National Laboratory, Los Alamos, New Mexico 87545, USA}

\author{Matthew T. Reiten}
\affiliation{Center for Integrated Nanotechnologies, Los Alamos National Laboratory, Los Alamos, New Mexico 87545, USA}

\author{Sheng-Nian Luo}
\affiliation{P-25, Los Alamos National Laboratory, Los Alamos, New Mexico 87545, USA}

\author{Abul K. Azad}
\affiliation{Center for Integrated Nanotechnologies, Los Alamos National Laboratory, Los Alamos, New Mexico 87545, USA}

\author{Antoinette J. Taylor}
\affiliation{Center for Integrated Nanotechnologies, Los Alamos National Laboratory, Los Alamos, New Mexico 87545, USA}

\author{Hou-Tong Chen}
\email[]{Author to whom correspondence should be addressed. Electronic mail: chenht@lanl.gov}
\affiliation{Center for Integrated Nanotechnologies, Los Alamos National Laboratory, Los Alamos, New Mexico 87545, USA}

\date{\today}

\begin{abstract}
We investigate the impact of resonator geometry and its coupling with ground plane on the performance of metamaterial perfect absorbers. Using a cross-resonator as an example structure, we find that the absorber thickness can be further reduced through modifying the geometric dimensions of the resonators. Numerical simulations and theoretical calculations reveal that destructive interference of multiple reflections is responsible for the near-unity absorption. The near-field coupling between the resonator array and ground plane can be significant. When this coupling is taken into account, the theoretical results calculated using the interference model are in excellent agreement with experiments and numerical simulations.
\end{abstract}

\pacs{}

\maketitle 


Near-unity absorption of electromagnetic waves (or light) has many important applications including, but not limited to, improving energy harvesting efficiency, reducing radar cross-section, and enhancing detection sensitivity. A classical electromagnetic absorber, known as Salisbury screen,\cite{Munk2000FSS} consists of a resistive sheet and a metal ground plane separated by a dielectric spacer. Besides the zero transmission because of the ground plane, its operational mechanism is an impedance matching to free space due to destructive interference in reflection, essentially the same as in quarter-wave antireflection coatings. The requirement of a quarter wavelength spacer thickness, however, is disadvantageous, particularly in the long electromagnetic wavelength regimes. It also causes single frequency (or narrow band) operation and strong incidence angle dependence. One solution to reducing the spacer thickness is through the use of high-impedance surfaces replacing the conventional electrical ground plane, which is of low-impedance, and working as an artificial magnetic conductor.\cite{Sievenpiper1999IEEE,Engheta2002IEEE} They can be easily fabricated for microwaves but are difficult to fabricate in the terahertz (THz) and optical regimes.

The great interest in ultra-thin electromagnetic absorbers is largely due to the experimental demonstration of the so-called metamaterial perfect absorbers in 2008.\cite{Landy2008PRL,Tao2008OE} The structure of metamaterial absorbers is in fact one kind of lossy high impedance surface that has been shown capable of high absorption.\cite{Kern2003} Employing multi-layered metamaterials or unit cells containing structures resonating at different frequencies, multiband or broadband metamaterial absorbers have also been demonstrated.\cite{Wen2009APL,Tao2010JPD,Liu2011PRL,Singh2011APL,Shen2011OE,Cui2012NL,Huang2012OL} An interference theory was recently developed and successfully explained the observed perfect absorption and anti-parallel surface currents in two metallic layers;\cite{Chen2012OE,Chen2010PRL} the latter has been used by many researchers as the evidence of a magnetic resonance. Based on this interference theory, any subwavelength resonator array on a dielectric layer backed with a ground plane can potentially serve as a high efficiency electromagnetic absorber, when the dielectric spacer thickness is appropriately chosen. In contrast to Salisbury screens which require a dielectric spacer of a quarter wavelength thickness to provide the necessary propagation phase and satisfy the destructive interference, such a thickness requirement is largely lifted thanks to the resonant response and strong dispersion of the metamaterial layer.\cite{Chen2012OE} It is the air-spacer interface with the planar resonator array that largely provides the phase shift that supports the destructive interference. 

In this work we investigate, both experimentally and numerically, the impact of resonator geometry on the performance of metamaterial perfect absorbers. Using an array of cross-resonators,\cite{Liu2010PRL, Grant2011OL} we find that the spacer thickness can be significantly reduced (by a factor of two) through increasing the wire width of the cross-resonators. Additionally, the results are in good agreement with theoretical results using the interference model for small wire width of the cross-resonators. However, significant deviation in peak absorption frequency is observed for increasing wire width of the cross-resonators, due to the capacitive coupling between the cross-resonators and the ground plane. We take a simulation strategy so that this coupling is taken into account in the interference calculations, and achieve excellent agreement.

\begin{figure}[t]
\centerline{\includegraphics[width=3 in]{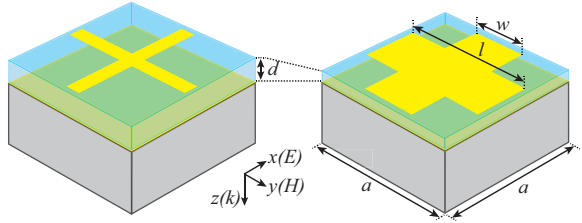}}
\caption{Unit cell geometry of the metamaterial perfect absorbers, where the wire width $w$ of the cross-resonators and the thickness $d$ of the dielectric spacer are varied.}\label{Fig1}
\end{figure}

The structures of the metamaterial absorbers employed in this investigation are shown in Fig.~\ref{Fig1}. The unit cell consists of a gold cross-resonator on a polyimide dielectric layer backed with a gold ground plane, where the symmetric structure is polarization insensitive for normal incidence. The thicknesses of the gold cross-resonator and ground plane are both 200~nm, while the polyimide spacer thickness $d$ is variable and ranges from 4~$\mu$m to 8~$\mu$m. The cross resonator has a periodicity $a = 85$~$\mu$m, an outer dimension $l = 80$~$\mu$m, and a variable wire width $w$ ranging from 10~$\mu$m to 80~$\mu$m. The numerical simulations were carried out using CST Microwave Studio 2011 and employing periodic boundary conditions. Gold was simulated as a Drude metal with a plasma frequency $f_p = 2181$~THz and a damping frequency $f_{\tau} = 6.5$~THz.\cite{Ordal1985AO} The polyimide spacer was treated as a dielectric material with a typical dielectric constant $\tilde \epsilon_s = 3(1+0.06 i)$. Under normal incidence the simulations provide S-parameters, and the absorption is obtained by $A = 1 - R - T = 1 - |S_{11}|^2$, where $A$ is the absorptance, $R = |S_{11}|^2$ is the reflectance, and the transmittance $T = 0$ due to the presence of the gold ground plane. The designs shown here are for operation near 1~THz.

The above metamaterial absorber designs were fabricated using standard micro-fabrication approaches. Briefly, the 10/200-nm-thick titanium/gold ground plane was first deposited on a substrate (here we used gallium arsenide) by e-beam evaporation, and then the polyimide spacer layer with varying thickness was spin-coated and thermally cured. The cross-resonator arrays were then fabricated using photolithographic methods, titanium/gold film deposition, and lift-off. The fabricated samples were characterized using a fiber-coupled THz time-domain spectrometer in the reflection mode, with an incidence angle of about $30^\circ$ due to its mechanical restriction.\cite{Huang2012OL,OHara2007JNO} A plain 10/200-nm-thick titanium/gold film was deposited on the identical substrate serving as the reference.

\begin{figure}[t]
\centerline{\includegraphics[width=3.4 in]{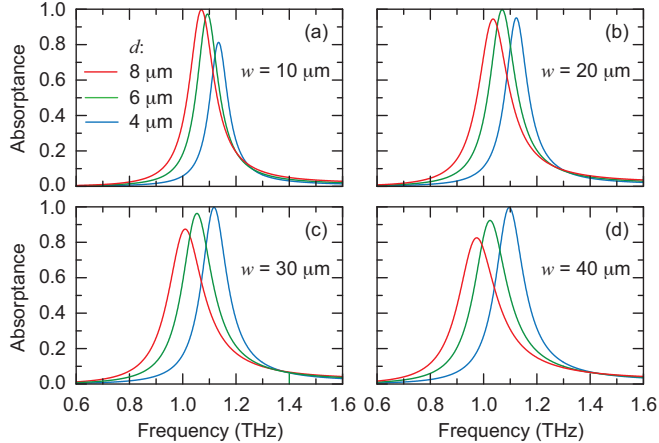}}
\caption{Numerically simulated absorption of the metamaterial absorbers with various wire widths of the cross-resonators and thicknesses of the polyimide spacer.}\label{Fig2}
\end{figure}
\begin{figure}[b]
\centerline{\includegraphics[width=3.4 in]{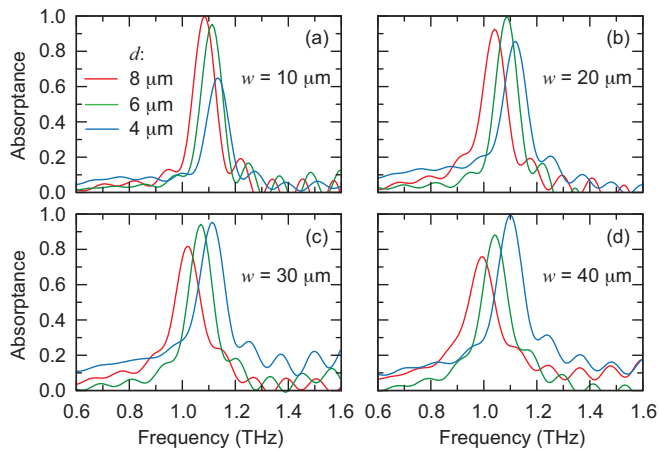}}
\caption{Experimentally measured absorption of the metamaterial absorbers with various wire widths of the cross-resonators and thicknesses of the polyimide spacer.}\label{Fig3}
\end{figure}

Numerical simulations were carried out for various wire widths of the cross-resonators and thicknesses of the polyimide spacer. The results are shown in Fig.~\ref{Fig2} for $w = 10, 20, 30, 40$~$\mu$m and $d = 4, 6, 8$~$\mu$m.  For a specific wire width $w$, there is always an optimized spacer thickness $d$ that can achieve near-unity absorption. For instance, for $w = 10$~$\mu$m, the optimized spacer thickness is about $d = 8$~$\mu$m. Deviating from this optimized thickness significantly decreases the absorption. More importantly, we find that the optimized spacer thickness decreases when the wire width of the cross-resonators increases. For $w = 20$~$\mu$m, the optimized spacer thickness decreases to about $d = 6$~$\mu$m, and for $w = 40$ $\mu$m, it further decreases to about $d = 4$~$\mu$m, only half of that for $w = 10$~$\mu$m. More accurate simulations show that the optimized spacer thicknesses are $d = 7.5$~$\mu$m and $3.6$~$\mu$m for $w = 10$~$\mu$m and $40$~$\mu$m, respectively. Additionally, although not shown, the numerical results reveal that increasing the wire width beyond $40$~$\mu$m has very little further impact on the optimized spacer thickness, as indicated by the minimal change when comparing Fig.~\ref{Fig2}(c) for $w = 30$~$\mu$m and Fig.~\ref{Fig2}(d) for $w = 40$~$\mu$m. The experimental results for the fabricated samples with the corresponding wire widths and spacer thicknesses are shown in Fig.~\ref{Fig3}, exhibiting excellent agreement as compared to the numerical results. That is, simply through varying the geometry of the resonators, we can reduce the overall thickness of metamaterial perfect absorbers from $\textit{d}/\lambda_0 = 1/36$ to $1/68$, where $\lambda_0$ is the wavelength in free space.

In contrast to the originally proposed impedance matching through a bulk effective media approach,\cite{Tao2008PRB,Hao2011PRB} the interference theory considers the cross-resonator array as a zero thickness, impedance tuned interface between the air (free space) and spacer dielectric.\cite{Chen2012OE} It modifies the transmission and reflection properties of the air-spacer interface, while the ground plane serves as a perfect reflector with a reflection coefficient of $-1$. Therefore the transmission through the metamaterial absorber is zero, and the overall reflection is a result of superposition of multiple reflections:
\begin{equation}
\tilde{r} = \tilde r_{12} - \frac{\tilde t_{12} \tilde t_{21} e^{i 2 \tilde\beta}}{1 + \tilde r_{21} e^{i 2\tilde\beta}} 
\label{Reflection}
\end{equation}
where $\tilde r_{12}$ and $\tilde r_{21}$, $\tilde t_{12}$ and $\tilde t_{21}$ are the reflection and transmission coefficients at the air-spacer interface with the cross-resonator array, incident from the air and spacer sides, respectively, and $\tilde \beta = \omega \sqrt{\tilde \epsilon_s} d / c_0$ is the complex propagation phase in the spacer, and $c_0$ is the light speed in free space. These coefficients can be obtained through numerically simulating S-parameters of a unit cell containing the cross-resonator located at the interface of semi-infinite air and spacer, i.e. without the ground plane and substrate. In this case we have assumed the near-field coupling is negligible between the resonator array and the ground plane, i.e., the ground plane does not affect the resonance of the cross-resonators.

\begin{figure}[htb]
\centerline{\includegraphics[width=3.4 in]{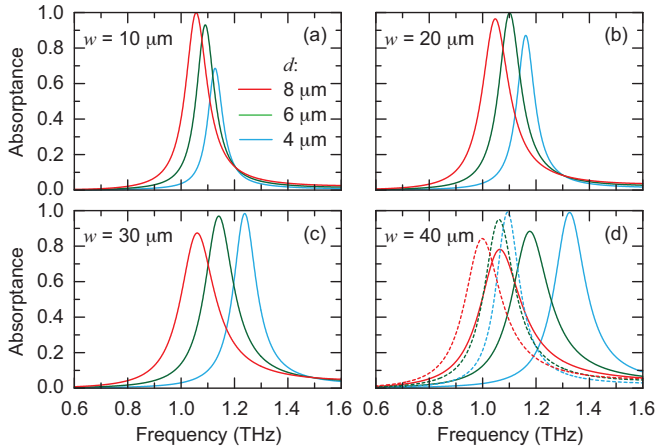}}
\caption{Analytically calculated absorption using the interference model for the metamaterial absorbers with different cross wire widths and spacer thicknesses. The dashed curves in (d) are for calculations that take into account the parasitic capacitance due to the presence of the ground plane.}\label{Fig4}
\end{figure}
The corresponding results calculated using the interference approach given by equation (\ref{Reflection}) are shown in Fig.~\ref{Fig4} for the above-mentioned wire widths and spacer thicknesses of the metamaterial absorbers. The resultant absorption spectra excellently reproduce the spacer thickness dependence on the wire width of the cross-resonators. For smaller wire width $w$, the calculated results also reproduce the peak absorption frequencies, which indicates the negligible near-field interaction. However, for larger wire width $w$, the calculated peak absorption frequency becomes significantly higher than the experimental and numerical values, suggesting an increasing effect of the ground plane on the resonance properties of the cross-resonator array. In fact, if the effect from the ground plane is not taken into account, the resonance frequency of the cross-resonator array increases with the wire width, which results in a higher value of the calculated peak absorption frequency.

\begin{figure}[htb]
\centerline{\includegraphics[width=2.5 in]{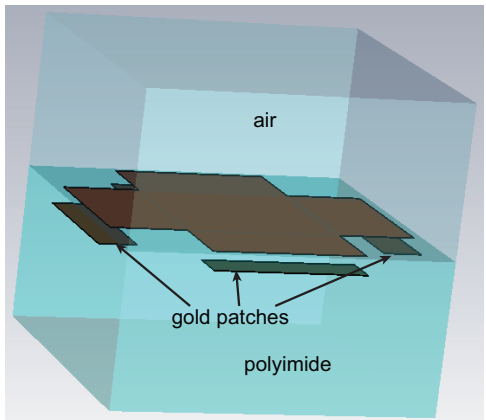}}
\caption{Unit cell used to simulate S-parameters at the interface of the cross-resonator array, where the parasitic capacitance from the ground plane are picked up by using the gold patches of 20~$\mu$m $\times$ 40~$\mu$m. Only half of each patch is shown in the unit cell. }\label{Fig5}
\end{figure}

Qualitatively, the presence of the ground plane provides additional capacitance that is in parallel with the inter-resonator capacitance. This reduces the resonance frequency of the cross-resonator array, and therefore the peak absorption frequency, as we observed in numerical and experimental results. Quantitatively, however, the presence of the ground plane prevents us from directly computing the resonant reflection and transmission coefficients at the interface of the cross-resonator array, which are required for the calculations using the interference model. Here we employ a simulation strategy of replacing the ground plane using gold patches underneath the gap between the neighboring resonators, as shown in Fig.~\ref{Fig5}. As a first order approximation, this takes into account the most important contribution of the ground plane, i.e. the parasitic capacitance, to the resonance of the cross-resonators. On the other hand, the relatively small dimensions (20~$\mu$m $\times$ 40~$\mu$m) of these patches result in negligible effects on the reflection and transmission, in contrast to the ground plane. The accordingly obtained reflection and transmission coefficients are then used in equation (\ref{Reflection}) to calculate the overall reflection and thereby the absorption. As shown by the dashed curves in Fig.~\ref{Fig4}(d), excellent agreement is achieved as compared to the numerical and experimental results shown respectively in Figs. \ref{Fig2}(d) and \ref{Fig3}(d).

In summary, we employ cross-resonator as a typical example of resonators for perfect metamaterial absorbers, and demonstrate numerically, experimentally and theoretically that the thickness of ultra-thin metamaterial perfect absorbers can be further reduced by modifying the geometric dimensions of the resonators. The results reveal that the zero reflection and near-unity absorption is due to a destructive interference of multiple reflections in the metamaterial absorber. For certain geometries and dimensions of the resonators, the ground plane can significantly affect the resonance frequency of the resonator array. If we take such an effect into account, the results from the theoretical interference model are in excellent agreement with experiments and numerical simulations.

Li Huang acknowledges support from Natural Science Foundation of China (NSFC) under Grant No. 10904023. We acknowledge partial support from the Los Alamos National Laboratory LDRD program. This work was performed, in part, at the Center for Integrated Nanotechnologies, a U.S. Department of Energy, Office of Basic Energy Sciences user facility.  Los Alamos National Laboratory, an affirmative action equal opportunity employer, is operated by Los Alamos National Security, LLC, for the National Nuclear Security Administration of the U.S. Department of Energy under contract DE-AC52-06NA25396.



%
%

%



\end{document}